# Spectral functions of CVD grown MoS2 monolayers after chemical transfer onto Au surface

Sung Won Jung,‡[a] Sangyeon Pak‡[b] Sanghyo Lee,*[c] Sonka Reimers,[a,d] Saumya Mukherjee, [a,e] Pavel Dudin,[a,h] Timur K. Kim,[a] Mattia Cattelan,[f] Neil Fox,[f,g] Sarnjeet S. Dhesi,[a] Cephise Cacho,*[a] and SeungNam Cha*[b]

[a] Diamond Light Source, Harwell Campus, Didcot OX11 0DE, United Kingdom

[b] Department of Physics, Sungkyunkwan University (SKKU), Suwon, Gyeonggi-do 16419, Republic of Korea

[c] Department of Engineering, University of Cambridge, 9 JJ Thomson Avenue, Cambridge CB3 OFA, United Kingdom

[d] School of Physics and Astronomy,The University of Nottingham, University Park, Nottingham NG2 7RD Nottingham, UK

[e] Clarendon Laboratory, Department of Physics, University of Oxford, Parks Road, Oxford OX1 3PU, United Kingdom

[f] School of Chemistry, University of Bristol, Cantocks Close, Bristol BS8 1TS, UK

[g] H. H. Wills Physics Laboratory, University of Bristol, Tyndall Avenue, Bristol BS8 1TL, UK

[h] Synchrotron SOLEIL, L'Orme des Merisiers, Saint-Aubin, 91190 Gif-sur-Yvette, France

‡ S. Jung and S. Pak contributed equally to this work.

* Corresponding author. Tel: +44 (0)1223 748318

E-mail address: cephise.cacho@diamond.ac.uk, chasn@skku.edu, sl920@cam.ac.uk

# Abstract

The recent rise of van der Waals (vdW) crystals has opened new prospects for studying versatile and exotic fundamental physics with future device applications such as twistronics. Even though the recent development on Angle-resolved photoemission spectroscopy (ARPES) with Nano-focusing optics, making clean surfaces and interfaces of chemically transferred crystals have been challenging to obtain high-resolution ARPES spectra. Here, we show that by employing nano-ARPES with submicron sized beam and polystyrene-assisted transfer followed by annealing process in ultra-high vacuum environment, remarkably clear ARPES spectral features such as spin-orbit splitting and band renormalization of CVD-grown, monolayered $MoS_2$ can be measured. Our finding paves a way to exploit chemically transferred crystals for measuring high-resolution ARPES spectra to observe exotic quasi-particles in vdW heterostructures.

Keywords: $MoS_2$ monolayer, nano-ARPES, chemical vapor deposition, polymer-assisted transfer, inelastic scattering

The recent rise of van der Waals crystals has opened new prospects for studying versatile and exotic fundamental physics,[1] as well as having immense future potential in functional devices due to their wide range and tunability of the electronic band structures associated with quantum confinement effects and interlayer coupling.[2-4] Especially, such well-ordered vdW crystals can be exfoliated to have atomic-scale thickness (Figure 1a), which have opened enormous potential for various device applications as well as novel scientific findings.[5-8] For example, bilayer graphene shows unconventional superconductivity with a magic twisted-angle,[9] and new excitonic behaviors were observed for transition metal dichalcogenides (TMDCs) bilayers at a slight twist angle.[10] To investigate various exotic quantum phenomena and novel devices composed of one or two monolayers thick crystals, the primary concern lies on uncovering the key aspects of the electronic band structures in momentum space, as their band structure possess information about the charge, spin, and electron that are endowed with a valley degree of freedom.

In this regard, angle-resolved photoemission spectroscopy (ARPES) is one of the few techniques that give direct access to the comprehensive electronic structure information of solid materials.[11] It directly measures escape angle and kinetic energy of the photoemitted electrons, so that electron bands of crystals could be reconstructed in momentum space. In the extreme ultraviolet (XUV) photon energy region (<100 eV), ARPES is surface sensitive strongly due to the short limited inelastic mean free path of electrons, making the technique ideal for investigating the band structure of two-dimensional vdW crystals.[11] More recently, there has been a development on sub-micron sized beam focusing with Fresnel zone-plate (FZP)[12] and capillary mirror,[13] which led to achieve in resolving electronic structure of few micron-sized vdW flakes and CVD-grown vdW crystals. Especially, such spatially-resolved ARPES technique was utilized to successfully probe the exotic features of vdW crystals, such as gate bias-dependent electronic structure,[14] one-dimensional edge states,[15] electronics structure between several domains.[16] Especially, as spatially-resolved ARPES is able to operate with inducing a bias on the crystal, the technique has garnered intensive interest regarding condensed matter physics as well as device applications with two dimensional materials.[14]

Although spatially resolved ARPES technique has been widely employed for resolving electronic structures, there still remain challenges for exploring atomically thin vdW materials and heterostructures which need to be prepared by chemical-based transfer method. That is, any surface chemical contamination arisen from transferring of exfoliated or CVD-grown vdW crystals would hinder to obtain clear ARPES spectra due to an increase of electron scattering at the surface or at the interface between the crystal and substrate.[17-20] Furthermore, low energy excitations in ARPES spectra of chemically transferred flakes has not been available to measure, although typical ARPES studies on TMDCs with the surface preparation in ultra-high vacuum (UHV) have obtained exotic spectral features with high-energy resolution, such as the band transition between direct- and indirect- bandgap,[21] electron-boson interaction,[22] anomalous polaron states,[23] and fingerprints of trions[24]. Therefore, it is desired to find a proper transfer and sample preparation method for clean surface and interface of vdW crystals to resolve highly clear ARPES spectra.

In this article, we show the highly-symmetry energy band structure of CVD-grown $MoS_2$ monolayer measured using spatially resolved ARPES (Figure 1b). The micro- crystallites of $MoS_2$ monolayer were grown by CVD method and were prepared on a Au substrate for the spatially resolved ARPES measurement. A suitable transferring and cleaning method are employed to minimize chemical contamination before ARPES measurement. Thermal treatment of samples in ultra-high vacuum (UHV) were employed to efficiently remove residual contamination after polystyrene (PS)-assisted transfer of $MoS_2$. Then, we could observe clear band spectra of $MoS_2$. The obtained band spectra show spin-orbit splitting of 149 meV at the valleys and momentum mediated band renormalization between $MoS_2$ and Au substrates. Our spectral analysis shows the quality of spectral function in chemically transferred crystal is comparable with $MoS_2$ monolayers grown on Au surface using epitaxial methods such as MBE in UHV environment.[25-27] Our finding paves a way to exploit chemically transferred TMDC crystals and other challenging materials for measuring band spectra and other exotic quasi-particles in vdW heterostructures and electronic/spintronic devices.

**Results and discussion**

CVD synthesis of monolayer $MoS_2$ was carried out at 800 ˚C for a duration of 5 minutes inside a 1-inch quartz tube furnace at atmospheric pressure. The schematic in Figure 2a describes the CVD setup and the arrangement of the $SiO_2$/Si substrate and the boat containing the $MoO_3$ precursor. As we previously reported,[28,29] 0.01 mg of $MoO_3$ precursors were used in order to grow $MoS_2$ monolayers in a thermodynamically stable reaction to occur during the growth processes. The resulting as-grown $MoS_2$ monolayer on the $SiO_2$/Si substrate is shown in Figure 2b. In order to verify that the synthesized $MoS_2$ crystals are monolayers, we conducted Raman and photoluminescence analysis. Figure 2c shows the two characteristic Raman peaks observed at 383 $cm^{-1}$ and 401 $cm^{-1}$, corresponding to the in-plane ($E^1_{2g}$) and out-of-plane ($A^1_g$) vibrational modes, respectively. The difference in the two modes is around 18 $cm^{-1}$, confirming that the synthesized $MoS_2$ is monolayered. We also performed PL analysis as shown in Figure 2d. The PL spectra show strong emission peak centered at 1.832 eV with its full width at half maximum (FWHM) of 53 meV. The strong emission, as well as small FWHM, corresponds to the direct band gap and high-crystallinity characteristic of MoS$_2$ monolayer.[28] The monolayer feature of MoS$_2$ was further identified on the SiO$_2$ substrate using atomic force microscopy. The thickness of the as-grown MoS$_2$ shows its thickness around 0.7 nm, corresponding to single-layered of MoS$_2$.

In order to employ the CVD grown MoS$_2$ monolayer in the spatially resolved ARPES measurement, the monolayers were transferred from the SiO$_2$/Si substrate onto an amorphous Au thin film with a thickness of 100 nm on SiO$_2$/Si substrate. Note that the Au surface is required for ARPES measurement to make contact path between MoS$_2$ crystals and the ground and avoid charging effects in photoemission. Moreover, observation of hybridization of electronic states between the MoS$_2$ and Au can be a clear signature of the cleanness of the interface between them.[25-27] The transfer process is described in Figure 2f. We spin coated polystyrene (PS) onto the sample substrate. The PS film holds MoS$_2$ crystals, while water penetrates between the substrate and PS film so that the PS/MoS$_2$ film can be detached from the substrate, the process known as the surface-energy-assisted transfer.[30] This transfer process can avoid undesired damages to the MoS$_2$ crystals caused by etching of growth substrate in a base solution bath (KOH or NaOH) or hydrogen bubble to lift off the film, as conventionally used to transfer 2D crystals (Figure S1).[30] The detached PS/MoS$_2$ film is then suspended and dried in the air to remove water molecule. The film is transferred onto the Au/SiO$_2$/Si substrate, and the PS film is removed by dipping the samples in toluene. The transferred MoS$_2$ monolayer can be seen in Figure 2f.

The transfer process is a chemical-based technique and leaves undesired chemical residues, which can play as disorders during the photoemission process, on the surface of the MoS$_2$ crystals. In order for further effective removal of organic residues on the surface of MoS$_2$ crystal, we conducted a post-annealing in UHV environment and measured the ARPES spectra because of high sensitivity of spectral function in ARPES that is related to homogeneities and impurities. The efficiency of post-annealing was monitored with a photoemission electron microscopy (PEEM) in Nano-electron spectroscopy for chemical analysis (Nano-ESCA) that is a convenient technique to measure both real space by PEEM and reciprocal space by full-wavevector ARPES from few micron-sized crystals.[31-33]

Figure 3a,b show energy-filtered PEEM images of $MoS_2$ monolayers transferred onto $Au/SiO_2/Si$ substrate with the post-annealing process at 350 ˚C and 400 ˚C for 12h.[34,35] The obtained PEEM images do not present any noticeable difference between the samples, implying that the $MoS_2$ crystals are stable without structural damages under the post annealing up to 400 ˚C (Figure S2), and the PS based surface-energy-assisted transfer method could be introduced for obtaining the clean surface of TMDC when compared to those of others.[34] However, the spatial resolution of the PEEM measurement is not enough to resolve nanometer-sized chemical contaminations on the surface. Therefore, we carried out the measurement of constant energy map on the monolayered $MoS_2$ single crystal utilizing full-wavevector ARPES mode in Nano-ESCA. We choose 1.6eV binding energy for constant energy map for ARPES mode due to its convenience for distinguishing the contrast between valance band maxima (VBM) of $MoS_2$ and background. Constant energy maps in Figure 3c,d show clear difference between samples post-annealed at 350 ˚C and 400 ˚C, even though the difference is not observable by PEEM. The constant energy map of $MoS_2$ annealed at 350 ˚C shows mostly single blurred spot at Γ point, and this spot is observed on a relatively high background. For the sample annealed at 400 ˚C, while, the hexagonal arrangement of dimmed spots at K, K' valleys become highly clear. These constant energy maps imply that chemical residues are effectively removed from the surface by annealing at 400 ˚C because the boiling point of PS could be shifted below 400 ˚C under the ultra high vacuum condition. The transferring technique we adopted with post-annealing process at 400 ˚C could give

diverse opportunities to measure the band structure of transferred CVD grown $MoS_2$ crystals by ARPES measurement.

To confirm the surface status further, in addition, we obtained Mo 3*d* and S 2*p* core-level spectra through PEEM at the end-station in Diamond Light Source (UK). We spanned energy-filtered PEEM images with fixed photon energy, we can obtain core-level spectra of $MoS_2$ monolayers. Photoemission spectrum of Mo 3*d* core-level in figure 3e shows a spin-orbit coupled doublet near S 2*s* core-level. Moreover, there are two types of spin-orbit coupled doublets in S 2*p* core-levels. Such two doublets can be interpreted by two sulfur layer, upper layer towards vacuum and lower layer at the interface between $MoS_2$ and Au. These core-level spectra are well consistent with previous report for epitaxially-grown $MoS_2$ monolayers on Au surface,[36] indicating that our approach to preparing samples is promising to make a clean surface enough to measure the spectral function of valance bands by a highly surface-sensitive technique such as ARPES.

In order to measure the clear spectral feature, we performed the measurement of ARPES spectra using Nano-ARPES end-station in Diamond Light Source (UK). Here the Fresnel zone plate (FZP) optics is used to focus light into spot with sub-micron sizes (750 nm). The energy resolution of the ARPES measurement is ~30 meV. Figure 4b shows the spatially scanned photoelectron intensity map of the monolayer $MoS_2$ crystal transferred on $Au/SiO_2/Si$ substrate. The photoemission map is consistent with the optical microscope image in Figure 4a and shows a $MoS_2$ monolayer with a triangular shape which is the typical shape of CVD grown $MoS_2$ crystals, indicating that a few microns-sized single crystals can be resolved clearly with the measurement set-up. As the triangular shape of $MoS_2$ crystals is driven by its atomic structure,[37] we can identify the ΓK direction from the photoelectron intensity map. The sample was rotated to align with analyzer slit along ΓK direction in order to measure ARPES spectra from Γ to K at once.

Figure 4c show ARPES spectra of a monolayer $MoS_2$ crystal on the $Au/SiO_2/Si$ substrate. It shows clear ARPES spectra of monolayer $MoS_2$ crystal with spin-orbit splitting feature at K valley. In order to examine the spin-orbit splitting feature in detail, the energy distribution curves (EDC) at K valley (white dotted line) were analyzed as shown in Figure 4d. The EDC shows perceptible band splitting of the valence band resulting from the spin-orbit coupling without any mathematical treatments, such as second derivative or curvature extracting technique,[17] which have been required for observing the splitting in the previous literature.[38] The EDC was also well reproduced by a double Voigt functions fitting with a polynomial backgrounds subtraction, and the band profile of two spin-split states was plotted in Figure 4d (red and blue curves). Note that disorders can increase the full-width at half-maximum (FWHM) of Lorentzian function of ARPES spectra, and by analyzing the FWHM of the spectra, we can investigate disorders which lead to weaker contrasts between dispersions and backgrounds. Table I shows the binding energy, FWHM of Voigt function, Gaussian and Lorentzian width of two spin-split states at K valley, indicating the spin-split feature of 149 meV which is consistent with previous ARPES study of $MoS_2$ monolayers grown by epitaxial growth method on Au (111) surface.[25,26] Especially, the Lorentzian width of upper-state is 68 meV. The sharp Lorentzian shape (68 meV) of the upper band implying that the quality of electrical band on chemically transferred CVD grown $MoS_2$ crystals is comparable with that of epitaxially grown crystals (about 50meV).[27] Also, the obtained effective mass at VBM (Figure 4e) is similar to the predicted effective mass for freestanding $MoS_2$,[19] implying the negligible interaction between $MoS_2$ and Au at K point, and this is consistent with general agreement regarding momentum dependent feature for interlayer coupling of $MoS_2$ layers.

In addition, we also focused on spectral features at Γ where the overlap of wavefunction between $MoS_2$ and Au is maximized. Figure 4c shows that the valence bands of $MoS_2$ crystal is overlapped with the bulk state of Au film near Γ, and there is 100 meV energetic mismatch between measured dispersion of $MoS_2$ monolayer and calculated dispersion of freestanding $MoS_2$ monolayer (black dashed line). Considerable energy difference between Γ and VBM is about 300meV (Figure S3). It is similar with the ARPES spectra of renormalized bands for $MoS_2$ on Au (111) surface,[26] confirming that the band hybridization at Γ point occurs at the interface between $MoS_2$ crystal and Au thin-film. The results of spectral function analysis indicate not only the homogeneity of system but also the reduction of chemical impurities at the surface and the interface between $MoS_2$ and substrate, which is achieved through post annealing under UHV condition. We note that our ARPES data shows dispersion of Au bands, which indicates the crystallization of Au underneath the MoS2 monolayers by annealing.

Our experimental approach using CVD-grown $MoS_2$ crystals offers a viable route to establish both the crystal orientation/direction and the direction of high-symmetry points in momentum space as they are determined by the triangular shape of CVD-grown $MoS_2$ (Figure 1a), as also seen through HR-TEM image of $MoS_2$. Such benefit leads to the ease of ARPES alignment as well as fabrication of vdW heterostructures, which has previously needed to be found beforehand through alternative measurement techniques such as Low Energy Electron Diffraction and constant energy map of photoelectrons.[37] In this regard, using CVD-grown TMDC crystals in an ARPES measurement provide a means for exploiting clear and distinguishable spectral functions of CVD-grown vdW heterostructure and especially twisted heterobilayers (also known as twistronics),[9,10,39,40] that are being extensively studied very recently. Our spatially-resolved ARPES study shown in Figure 1b demonstrates the successful utilization of the CVD-grown $MoS_2$ crystal prepared through unique transfer method/substrate condition and measurement setup. Therefore, our approach offers a potential procedure for acquiring highly clear ARPES spectra with known crystal orientation.

**Conclusions**

In conclusion, we have shown an effective method for the measurement of highly defined ARPES spectra features using chemically transferred CVD-grown $MoS_2$ monolayer. This is demonstrated by employing a proper transferring method that reduces transfer-induced damages and chemical residues on the crystals, and a post-annealing treatment at 400 ˚C that is effective in removing unintentional chemical residues that disturb the acquisition of surface and interface sensitive ARPES spectra. The reduction in disorders in ARPES spectra were confirmed using PEEM technique on a laboratory-based Nano-ESCAII system. Using spatially-resolved ARPES technique with sub-micron sized beam, we succeeded in measuring ARPES spectra of chemically transferred CVD-grown $MoS_2$ monolayer with 149 meV spin-split at valleys and momentum mediated renormalization with Au substrate.[25-27] Our results confirm the feasibility of CVD-grown TMDC crystals with an effective sample preparation procedure which opens up the opportunities to explore the electronic bands in chemically fabricated samples for electronic-, spintronic- and twistronic- devices, as well as exotic quai-particles such as polarons and trions.

**Acknowledgements**

This work was carried out with the support of the Diamond Light Source, I05 and I06 beamline (proposal NT22901 and MM24367) and was supported from the National Research Foundation (NRF) of Korea (2019R1A2C1005930). The authors also acknowledge the Bristol NanoESCA Facility (EPSRC Strategic Equipment Grant EP/K035746/1 and EP/M000605/1) and the European Unions Horizon 2020 research and innovation programme (Marie Skodowska-Curie Grant Agreement 665593).

**Table 1**. Characterized parameter of spin-split pairs at K valley. All parameters were fitted by using Voigt function, convoluted function with Gaussian and Lorentzian function.

| | $E_B$ at K (eV) | | FWHM | Gaussian width (eV) | | Lorentzian width (eV) | | $m_{eff}$ (m/$m_0$) | |
|---|---|---|---|---|---|---|---|---|---|
| Upper | -1.388 | ±0.002 | 0.089 | 0.044 | ±0.004 | 0.068 | ±0.008 | 0.57 | ±0.03 |
| Lower | -1.537 | ±0.003 | 0.136 | 0.066 | ±0.005 | 0.104 | ±0.006 | 0.84 | ±0.08 |

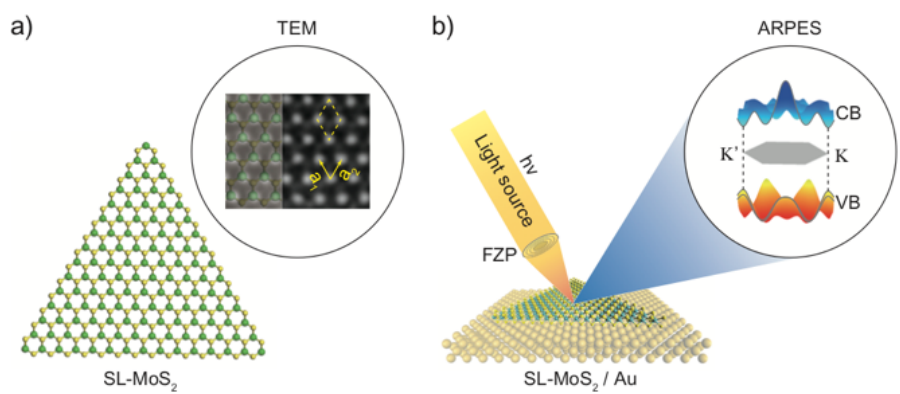

**Figure 1.** (a) Schematic atomic structure of the CVD grown crystal. (b) Schematics for the concept of Nano-ARPES end-station in Diamond Light Source. The incident beam was focused by FZP optics, is about 750nm, and beam spot size is about 750nm.

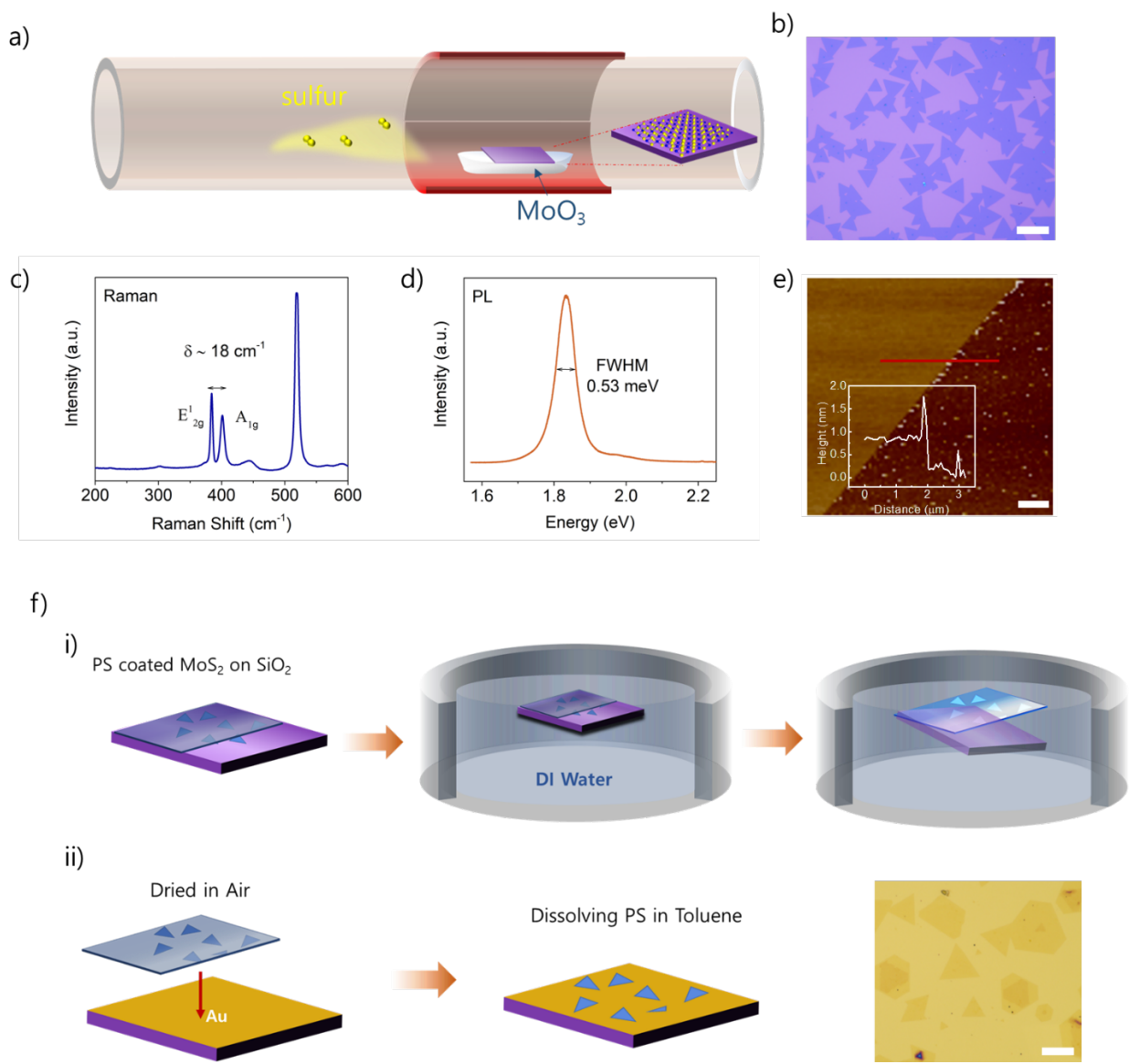

**Figure 2.** MoS$_2$ monolayer sample preparation for ARPES measurement. (a) Illustration of the CVD synthesis process for a MoS$_2$ monolayer. (b) Optical image of the as-grown MoS$_2$ monolayers. Scale bar: 50 µm. (c) Raman spectrum of the MoS$_2$ monolayer. (d) PL spectrum of the MoS$_2$ monolayer showing strong emission centered at 1.832 eV. (e) AFM topography image of the MoS$_2$ monolayer. Scale bar: 1 µm (f) Schematic illustrations of MoS$_2$ transfer from the SiO$_2$/Si substrate to the Au/SiO$_2$/Si substrate. PS coated MoS$_2$ crystals are detached from the substrate by surface-energy-assisted transfer onto the Au/SiO$_2$/Si substrate. Scale bar: 50 μm

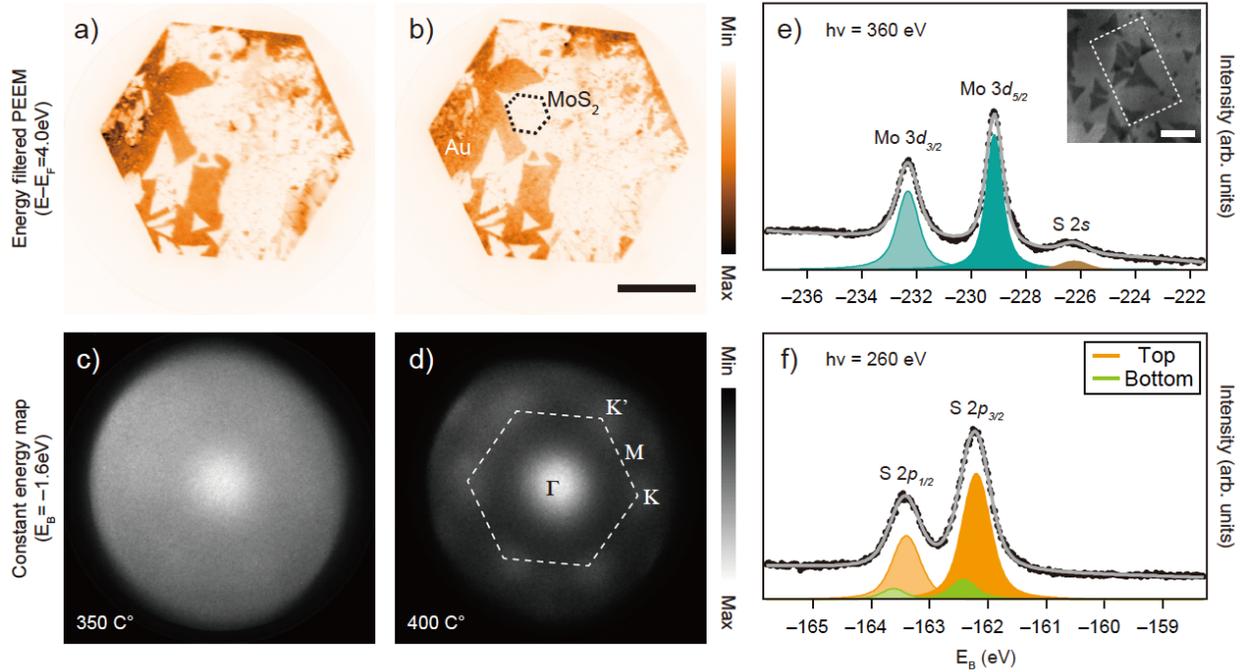

**Figure 3.** (a) and (b) Energy filtered PEEM image of MoS$_2$/Au/SiO$_2$ surface after annealing at (a) 350 ˚C and (b) 400 ˚C. (c) and (d) Constant energy maps of MoS$_2$ monolayer after anneal at (c) 350 ˚C and (d) 400 ˚C. Black dashed line in (b) show the area where constant energy map was taken. The shape of Brillouin zone of MoS$_2$ monolayer is outlined in (d) with white dashed line. A scale bar in (b) is 50 µm. Helium discharge lamp ($h\nu$=21.2eV) was used for excitation of photoemission in Nano-ESCA study. X-ray photoemission spectra of (e) Mo 3$d$ and (f) S 2$p$ core-levels, obtained by spanning energy filtered PEEM. Inset of (e) shows an image of the detection area of x-ray absorption, we reduced a size of beam-spot marked as a dashed line to obtain core-level spectra. Here the scale bar is 5 µm.

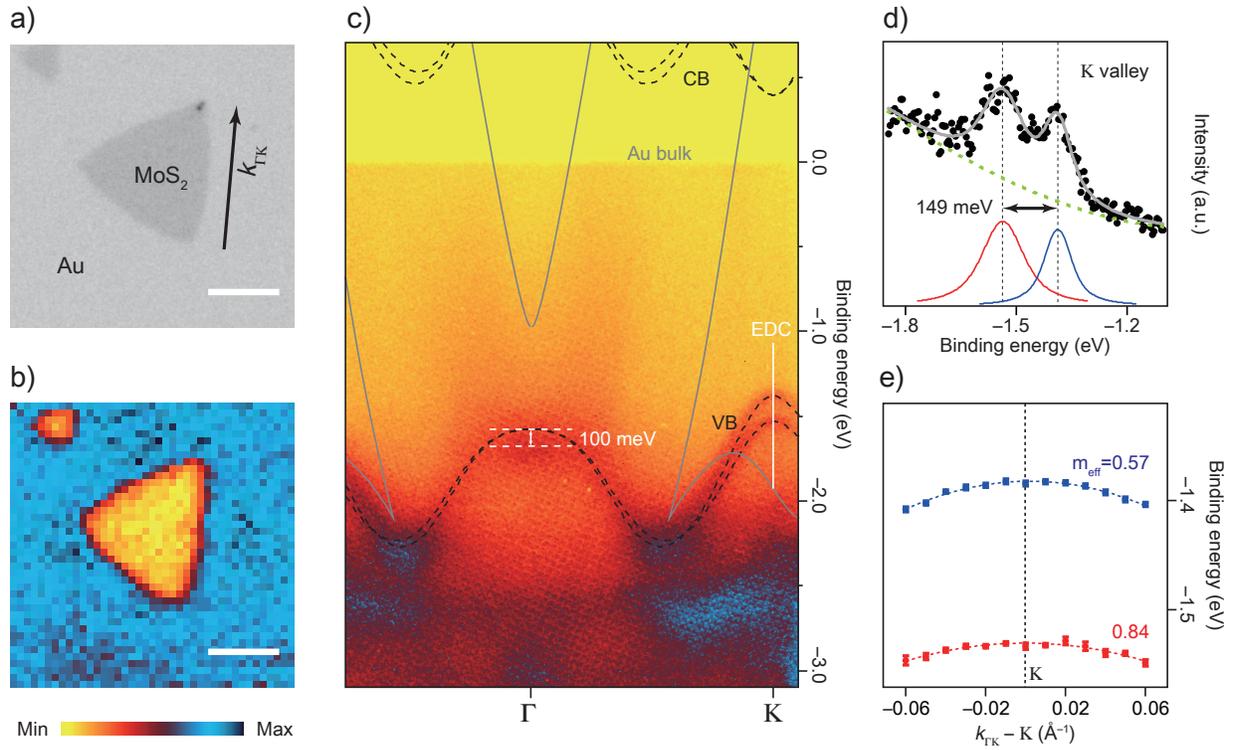

**Figure 4.** (a) Optical microscope image of MoS$_2$ monolayer transferred on Au surface. A white bar in (a) shows the length of 5$\mu$m. (b) Two-dimensional spatial map of (a). (c) ARPES spectra of MoS$_2$ monolayer (T=35K and $h\nu$=70eV) in (b). (d) Energy distribution curve (EDC) at K point in (c). (e) dispersion of two spin-split bands near K. Grey solid lines in (c) is the bulk projected bands, taken by ref.[41], black dashed lines in (c) are the electronic states of freestanding MoS$_2$ monolayer calculated by tight-binding method.[42] Binding energy of bands from calculation were shifted to match with valance band maxima (VBM) of MoS$_2$ monolayers taken by ARPES.

# Supporting Information for

# Spectral functions of CVD grown MoS2 monolayers after chemical transfer onto Au surface

Sung Won Jung,‡[a] Sangyeon Pak‡[b] Sanghyo Lee,*[c] Sonka Reimers,[a,d] Saumya Mukherjee,[a,e] Pavel Dudin,[a] Timur K. Kim,[a] Mattia Cattelan,[f] Neil Fox,[f,g] Sarnjeet S. Dhesi,[a] Cephise Cacho,*[a] and SeungNam Cha*[b]

[a] Diamond Light Source, Harwell Campus, Didcot OX11 0DE, United Kingdom

[b] Department of Physics, Sungkyunkwan University (SKKU), Suwon, Gyeonggi-do 16419, Republic of Korea

[c] Department of Engineering, University of Cambridge, 9 JJ Thomson Avenue, Cambridge CB3 0FA, United Kingdom

[d] School of Physics and Astronomy, The University of Nottingham, University Park, Nottingham NG2 7RD Nottingham, UK

[e] Clarendon Laboratory, Department of Physics, University of Oxford, Parks Road, Oxford OX1 3PU, United Kingdom

[f] School of Chemistry, University of Bristol, Cantocks Close, Bristol BS8 1TS, UK

[g] H. H. Wills Physics Laboratory, University of Bristol, Tyndall Avenue, Bristol BS8 1TL, UK

# S. Jung and S. Pak contributed equally to this work.

* Corresponding author. Tel: +44-(0)1223 748318

E-mail address: cephise.cacho@diamond.ac.uk, chasn@skku.edu, sl920@cam.ac.uk

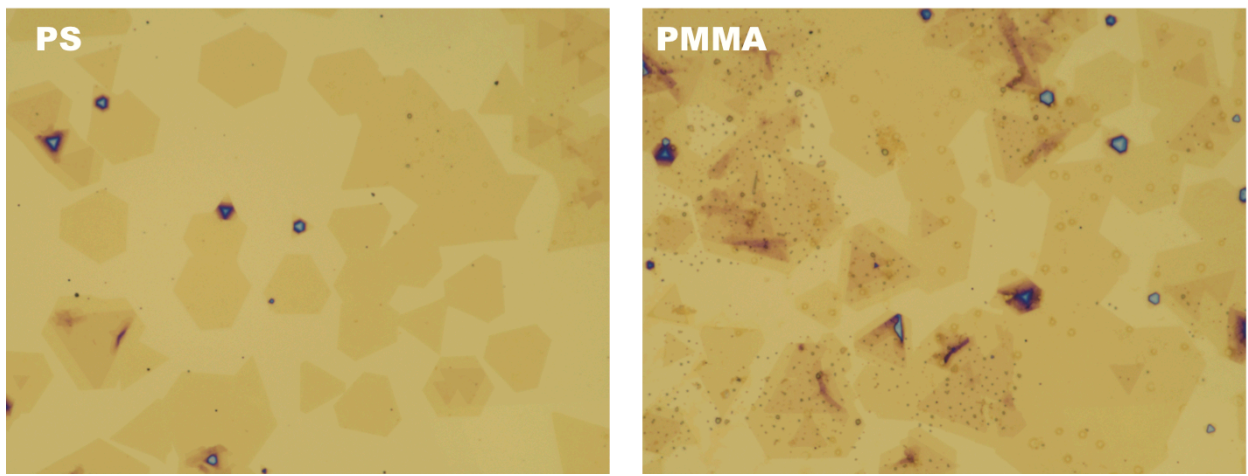

**Figure S1**. Optical image of MoS$_2$ crystals transferred onto Au/SiO$_2$/Si substrate using PS film (left) and PMMA film (right), showing that the PS-assisted transfer leads to clean surfaces.

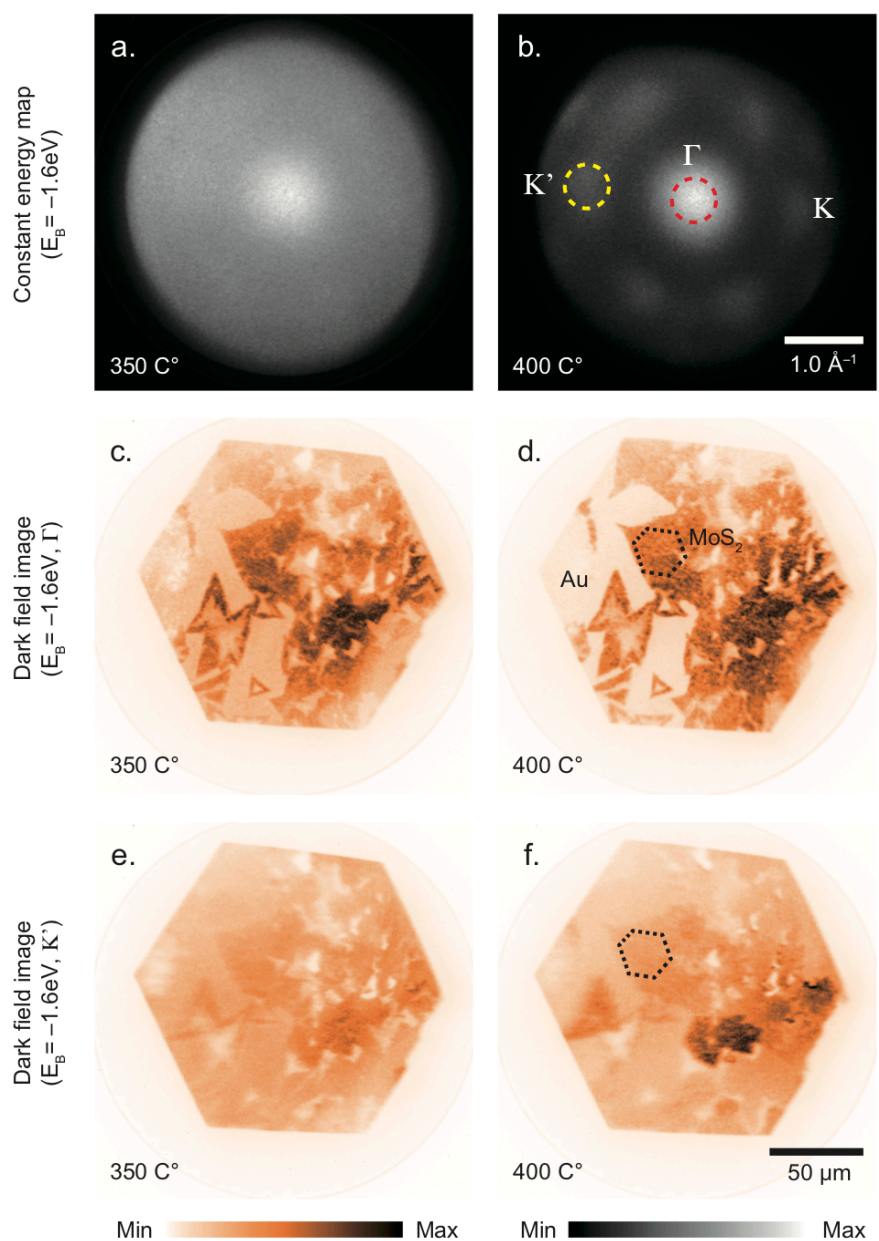

**Figure S2**. Constant energy map of MoS$_2$ monolayer after post-annealing at (a) 350 and (b) 400 °C for 12 hours. Dark field images filtered by the momentum space at (c, d) Γ and (e, f) K' on the samples after post-annealing at 350 °C and 400 °C, respectively. The obtained dark filed images show no degradation on the crystal shape after post-annealing at 400 °C even though the contrast of dark field images became higher.

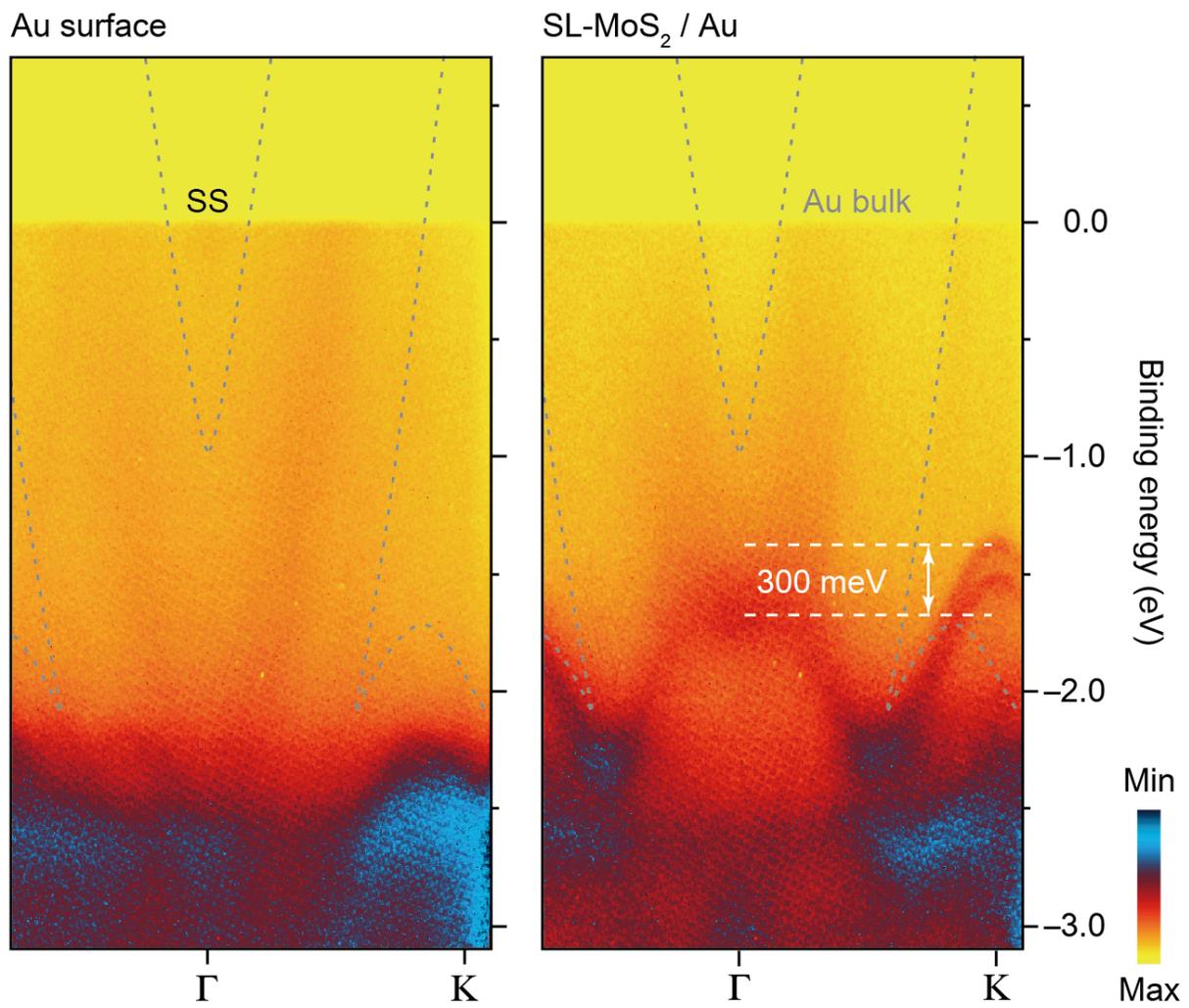

**Figure S3**. ARPES spectra of (left) Au surface and (Right) MoS$_2$ monolayer in figure 3b.

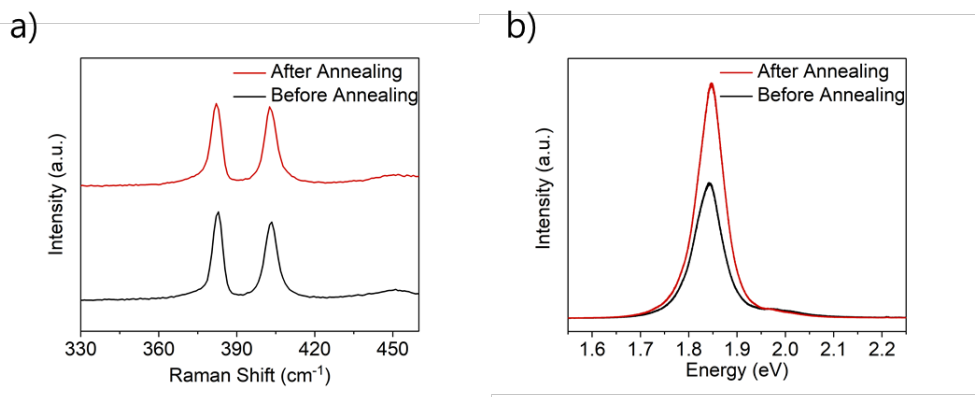

**Figure S4.** a) Raman and b) PL spectrum of transferred MoS$_2$ before and after UHV annealing at 400 ˚C.